\documentclass[twocolumn,aps,prl,superscriptaddress,longbibliography]{revtex4-1}
\usepackage[colorlinks,bookmarks=false,citecolor=blue,linkcolor=red,urlcolor=blue]{hyperref}
\usepackage{verbatim}
\usepackage{color}
\usepackage{amsmath,amssymb,bm,braket,float,mathtools,subfigure}
\usepackage{graphicx,xfrac,appendix}
\usepackage[english]{babel}

\makeatletter

\usepackage{epstopdf,float}

\makeatother

\begin{document}

\title{When Homogeneous Systems Meet Dissipation and Disorder}

\author{Xixi Feng}
\affiliation{Department of Physics, Zhejiang Normal University, Jinhua 321004, China}

\author{Ao Zhou}
\affiliation{Department of Physics, Zhejiang Normal University, Jinhua 321004, China}

\author{Feng Lu}
\affiliation{Department of Physics, Zhejiang Normal University, Jinhua 321004, China}

\author{Gao Xianlong}
\affiliation{Department of Physics, Zhejiang Normal University, Jinhua 321004, China}

\author{Shujie Cheng}
\thanks{chengsj@zjnu.edu.cn}
\affiliation{Xingzhi College, Zhejiang Normal University, Lanxi 321100, China}
\affiliation{Department of Physics, Zhejiang Normal University, Jinhua 321004, China}

\date{\today}

\begin{abstract}
We investigate the localization and topological properties of the non-equilibrium steady state (NESS) in a one-dimensional homogeneous system. 
Our results demonstrate that, despite the absence of disorder in the initial system, the NESS can exhibit localization under bond dissipation. 
These dissipation-driven localization and delocalization phenomena are clearly distinguished using Wigner distributions. 
Furthermore, we find that the initial localization characteristics of the system significantly influence the localization properties of the NESS. 
Drawing upon the concept of Bose-Einstein condensate broadening in cold-atom experiments, along with experimental data, 
we systematically characterize the impact of bond dissipation and disorder on the localization and topological properties of the NESS. 
The phase diagram reveals that the NESS can be topologically non-trivial even when the initial system is topologically trivial, and that the topological 
triviality of the initial system strengthens the topological non-triviality of the NESS. This work provides new insights into the 
localization and topological phase transitions in homogeneous systems induced by bond dissipation and disorder.

\end{abstract}

\maketitle

\section{Introduction}

Anderson localization theory occupies a central role in condensed-matter physics\cite{1958anderson}. 
Its contributions to the understanding of electron transport properties are profound. Based on scaling theory, 
a metal-insulator transition occurs in three-dimensional systems\cite{1979anderson}, separating an extended phase, 
where electrons propagate freely, from a localized phase, where electron motion is spatially confined. Localization 
phenomena have been observed experimentally across a diverse range of platforms. In cold-atom systems, precise 
control of interatomic interactions and external potentials allows for the creation of systems with tunable disorder, 
enabling clear observation of the transition from delocalized to localized states \cite{Billy2008,roati2008anderson,Modugno2010anderson,exp_4,exp_5,exp_6,exp_7,exp_8}. Furthermore, 
localization has also been observed in photonic quasicrystals \cite{lahini2009observation,dal2003light,verbin2013observation}.

In recent years, owing to the remarkable advancements in non-Hermitian physics \cite{NH_1,NH_2,NH_3,NH_4,NH_5,NH_6,NH_7}
and the precise regulation of dissipation and quantum coherence in experimental settings \cite{exp_NH1,exp_NH2,exp_NH3,exp_NH4},
the scientific community has witnessed a surging interest in dissipative open quantum systems. Dissipation has been shown to 
induce both localized and delocalized states, providing a key mechanism for understanding electron transport properties in 
disordered and homogeneous materials. \cite{dis_1,dis_2,dis_3,Longhi,Wang,Wang_2}.
Among them, it is known that the dissipation effects can not only disrupt localization and enhance transport \cite{dis_1,dis_2,dis_3},
but also induce mobility edges in systems that previously did not have delocalized-localized transitions \cite{Longhi}.
Moreover, the dissipation can drive the quasiperiodic system into specific states, which can be either delocalized or 
localized \cite{Wang,PRB_wang,Wang_3},
and regulate the topological properties of topologically nontrivial insulators so that they become
topologically trivial  or remain topologically nontrivial under steady-state \cite{Wang_2}.

In this work, we investigate previously unexplored aspects of dissipation-induced localization and 
topological phenomena. It is established that the NESS in bond-dissipative quasi-disordered systems 
is localized or delocalized, independent of the initial state\cite{Wang,bond_1,bond_2,bond_3,bond_4,bond_5,bond_6}. 
Here, we identify potential exceptions where NESS localization does depend on the initial localization and topological properties, 
possibly precluding localization entirely. Furthermore, we characterize the unexplored properties of the 
NESS in topologically trivial systems subject to both quasi-disorder and bond dissipation.
This investigation contributes to a deeper understanding of electron transport properties in 
homogeneous media subject to both bond dissipation and quasidisorder.

\begin{figure}[htp]
		\centering
		\includegraphics[width=0.5\textwidth]{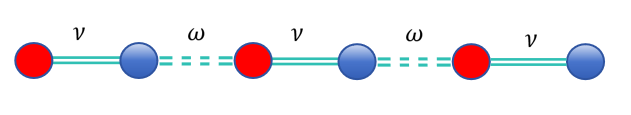}
		\caption{(Color online) Schematic diagram of the one-dimensional homogeneous system.
		$\nu$ and $\omega$ are hopping strength between neighboring sites. When $\nu \neq \omega$, 
		the system describe the two-band Su-Schrieffer-Heeger (SSH) model. When $\nu=\omega$, 
		the system describes the single band metal.
		}
		\label{f0}
\end{figure}

The paper is organized as follows. In Sec.~\ref{S1}, we briefly introduce the system and its Hamiltonian.
In Sec.~\ref{S2}, we reveal the localization and topological properties of the steady-state from the 
aspects of steady-state distribution and Wigner distribution. In Sec.~\ref{S3}, we study the effect of quasidisorder on localization
and topological property of the steady-state.
A summary is made in Sec.~\ref{S4}.

\section{Model and Band structures}\label{S1}

\begin{figure}[htp]
		\centering
		\includegraphics[width=0.5\textwidth]{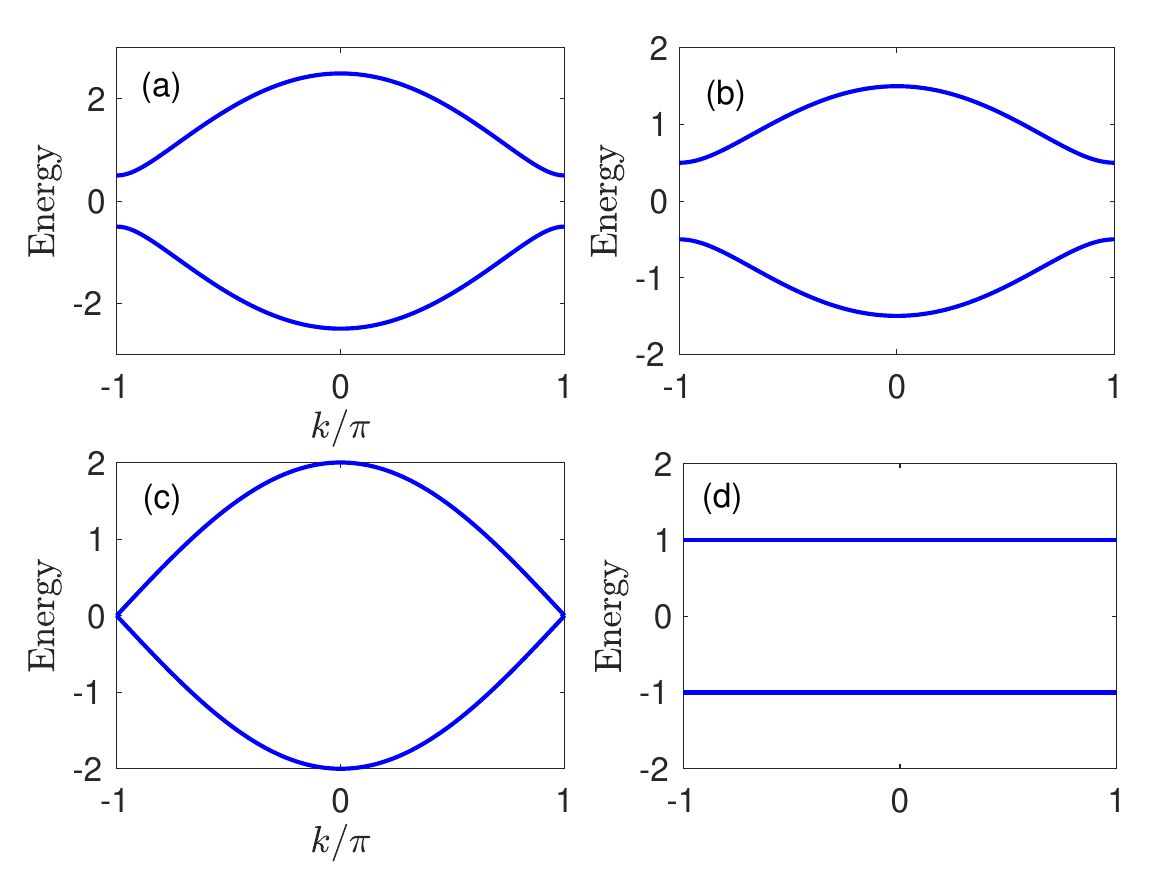}
		\caption{(Color online) Band structures of the one-dimensional homogeneous system.
		(a) $\omega=1.5\nu$; (b) $\omega=0.5\nu$; (c) $\omega=1\nu$ and (d) $\omega=0$.
		}
		\label{f1}
\end{figure}

We consider a one-dimensional homogeneous system (see the sketch in Fig.~\ref{f0}), whose
Hamiltonian is denoted as
\begin{equation}
H=\sum^{L-1}_{n=1,3,5,\cdots} \left(\nu c^{\dag}_{n+1}c_{n} +
\omega c^{\dag}_{n+2}c_{n+1}+{\rm H.c.} \right),
\end{equation}
where $\nu$ and $\omega$ are the hopping strengths between nearest-neighboring sites, and
$\nu$ is set as the unit of energy. By tuning the hopping strengths, leading to $\nu \neq \omega$,
the system becomes the two-band Su-Schrieffer-Heeger (SSH) model (see the band structures in
Figs.~\ref{f1}(a), \ref{f1}(b), and \ref{f1}(d)). It is known that when $\nu < \omega$,
the system under half filling behaves as a topological insulator, while it behaves normal insulator
when $\nu > \omega$ (Particularly, when $\omega=0$, the system is in the strongly insulating case with flat bands).
When $\nu=\omega$, the system describes a simplest single-band metal (see the band structure in Fig.~\ref{f1}(c)).
In an earlier cold-atom optical lattice experiment, Atala et al. proposed that the SSH model with or without energy
offset can be generated by two standing lasers with different wavelengths, and used an experimental technique of
coherent Bloch oscillations combined with Ramsey interference to measure geometric phases to
characterize the topological phase transition \cite{SSH_exp}. For single-band metal, it can be simulated by a similar way,
that is, a single-beam standing wave laser is used to construct a lattice potential well,
and the particle hopping between adjacent lattice points is realized by laser pumping \cite{roati2008anderson}.

\begin{figure}[htp]
		\centering
		\includegraphics[width=0.5\textwidth]{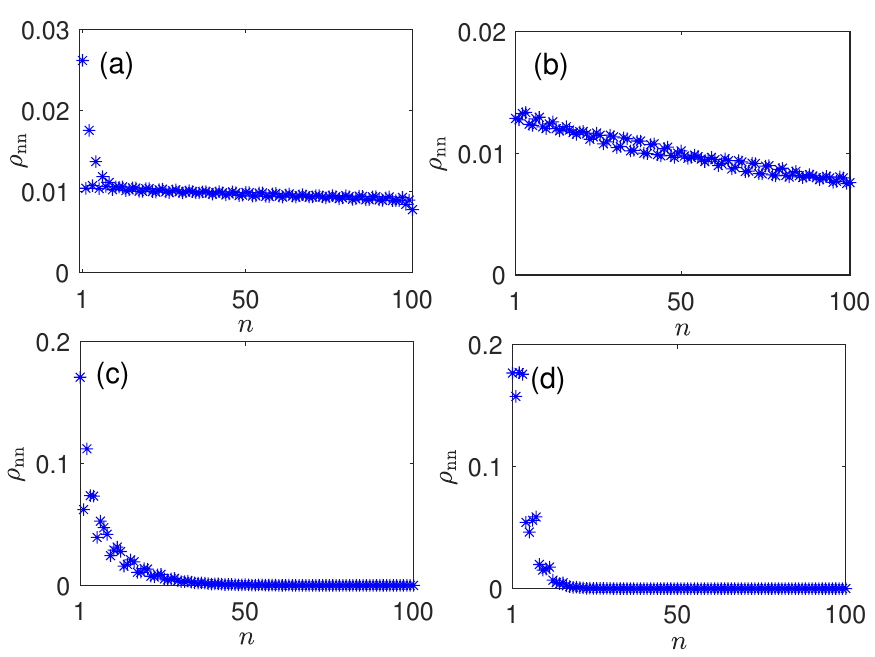}
		\caption{(Color online) Steady-state populations $\rho_{nn}$ as functions of site index.
		(a) $\gamma=0.1\nu$ and $\omega=1.5\nu$; (b) $\gamma=0.1\nu$ and $\omega=0.5\nu$; 
		(c) $\gamma=\nu$ and $\omega=1.5\nu$; (d) $\gamma=\nu$ and $\omega=0.5\nu$.
		Other parameters are $\ell=1$ and $L=100$.
		}
		\label{f2}
\end{figure}

\section{Non-equilibrium Steady-State, topological phase transition, and Wigner Distribution}\label{S2}
Previous work has shown that bond dissipation can induce a delocalized-to-localized state transition in 
quasiperiodic mosaic model \cite{Wang}. Observing that these dissipation-induced localized states exhibit 
edge localization characteristics, we are motivated to investigate whether bond dissipation can induce 
edge states in topological insulators even if the system is initially in the topological trivial phase. 
The relevant bond dissipation Lindblad operator $L_{n}$, constructed following Refs.\cite{bond_1,bond_2,bond_3,bond_4}, 
is given by 
\begin{equation}
L_{n}=\cos(2\pi\alpha n)c^{\dag}_{n}c_{n+\ell},
\end{equation}
 where the dissipation acts on a pair of neighboring sites $n$ and $n+\ell$,  
 and $\alpha=(\sqrt{5}-1)/2$ is an incommensurate parameter. We here briefly 
 introduce why we choose such type of bond dissipation. The bond dissipation operator comprises four 
 distinct contributions: two local dephasing terms $c^{\dag}_{n}c_{n}$ and $c^{\dag}_{n+\ell}c_{n+\ell}$ \cite{Longhi} 
 and and two non-local hopping terms $c^{\dag}_{n}c_{n+\ell}$ and $c^{\dag}_{n+\ell}c_{n}$. 
 Crucially, as established in Ref. \cite{Wang}, the dephasing components cannot drive a delocalization-localization 
 transition in the NESS. This stands in contrast to quasiperiodic modulations induced by incommensurate parameters, 
 which are rigorously proven both theoretically and experimentally \cite{roati2008anderson} to trigger such phase transitions.
 Therefore, we adopted the form of a combination of quasiperiodic modulation and transition operators as the bond dissipation operator, 
 in the hope of discovering some physical phenomena that are different from previous studies. More importantly, 
 this bond dissipation operator can be experimentally realized by introducing the auxiliary lattice \cite{auxiliary_1,auxiliary_2,auxiliary_3,auxiliary_4,auxiliary_5}. 
 At the end of the paper, we will address the experimental implementation of the proposed bond dissipation mechanism.
 After introducing such bond dissipation operator, then, the dynamical evolution of the density matrix $\rho$ is 
 governed by the following Lindblad master equation
\begin{equation}
\dot{\rho}=-i\left[H,\rho\right]+\gamma\sum_{n}\mathcal{D}\left[L_{n}\right]\rho,
\end{equation}
where the dissipator $\mathcal{D}\left[L_{n}\right]=L_{n}\rho L^{\dag}_{n}-\{L^{\dag}_{n}L_{n},\rho\}/2$, 
and $\gamma$ is the tunable bond dissipation strength.

The above evolution equation has the following equivalent expression
\begin{equation}
\rho(t)=e^{\mathcal{L}t}\rho(t=0),
\end{equation}
which contains all information of the system during the evolution. Due to the real parts of eigenvalues of the
matrix $\mathcal{L}$ ($\mathcal{L}$ is the Lindbladian matrix.) is non-positive, in the long-time evolution limit, the density matrix will eventually relax
to the zero-energy eigenstate of the matrix $\mathcal{L}$, i.e., the NESS $\rho_{ss}$. 

\begin{figure}[htp]
		\centering
		\includegraphics[width=0.5\textwidth]{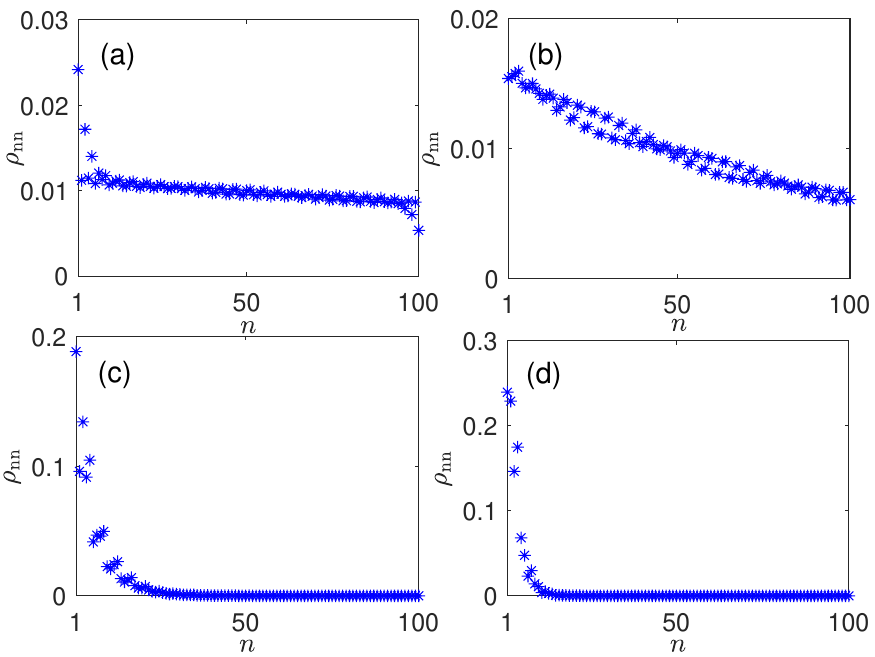}
		\caption{(Color online) Steady-state populations $\rho_{nn}$ as functions of site index.
		(a) $\gamma=0.1\nu$ and $\omega=1.5\nu$; (b) $\gamma=0.1\nu$ and $\omega=0.5\nu$; 
		(c) $\gamma=\nu$ and $\omega=1.5\nu$; (d) $\gamma=\nu$ and $\omega=0.5\nu$.
		Other parameters are $\ell=2$ and $L=100$.
		}
		\label{f3}
\end{figure}

We take open boundary condition in the following calculations. Under different parameters $\omega$ and $\gamma$, 
we obtain the corresponding steady-state density matrix $\rho_{ss}$, then the corresponding steady-state populations $\rho_{nn}$,
i.e., $\rho_{nn}=\langle n|\rho_{ss}|n\rangle$ (where $|n\rangle$ is the site basis) are calculated. Thus, the localization and 
topological properties of NESS can be characterized by the distributions of $\rho_{nn}$. 

We firstly consider the case under $\ell=1$ and $\gamma=0.1\nu$ and the system size is fixed at $L=100$ (the same below). 
Taking $\omega=1.5\nu$ and $\omega=0.5\nu$, the corresponding steady-state populations $\rho_{nn}$ 
are illustrated in Figs.~\ref{f2}(a) and \ref{f2}(b), respectively. We can see that under small dissipation strength, 
regardless of whether the initial system has topological nontrivial properties or not, it NESS is an extended state. It means that 
the weak bond dissipation can destroy the topological nontrivial property of the system.  
With larger dissipation strength $\gamma=\nu$, the corresponding $\rho_{nn}$ under $\omega=1.5\nu$ and $\omega=0.5\nu$ 
are plotted in Figs.~\ref{f2}(c) and \ref{f2}(d), respectively. We can see from Fig.~\ref{f2}(c) that the strong bond dissipation maintains 
the initially topological nontrivial properties because the NESS exhibits a localized edge state. As illustrated before, the system 
is in the topological trivial phase when $\omega=0.5\nu$. However, we can see from Fig.~\ref{f2}(d) that the NESS is an edge state. It means that 
the strong bond dissipation is conductive to the existence of topological edge states and it is feasible to prepare a topological nontrivial 
system by tuning the bond dissipation strength to a large value.

Similar circumstances happen at the $\ell=2$ case as well. Under $\gamma=0.1\nu$, the steady-state populations $\rho_{nn}$ 
for $\omega=1.5\nu$ and $\omega=0.5\nu$ are plotted in Figs.~\ref{f3}(a) and \ref{f3}(b), respectively. As seen that, the NESS 
for both cases are delocalized states. It means that the weak bond dissipation with $\ell=2$ can destroy the topological nontrivial 
property of the system, too. For larger dissipation strength $\gamma=\nu$, the corresponding $\rho_{nn}$ for $\omega=1.5\nu$ 
and $\omega=0.5\nu$ are plotted in Figs.~\ref{f3}(c) and \ref{f3}(d), respectively. It is seen that the distributions of NESS are both 
edge states. The results reflect the fact that the strong dissipation is not only conductive to the existence of topological edge state, 
and it is can be used to induce the new topological edge state. Therefore, if one wants to prepare a open topological nontrivial system, 
it is achievable to tune the bond dissipation strength to a large value.

Comparison of Figs.~\ref{f3} and \ref{f2} reveals that delocalization and edge localization in the NESS 
exhibit minimal dependence on $\omega$ and $\ell$, while demonstrating pronounced sensitivity to the dissipation strength $\gamma$.
The edge localization of the steady state reflects the system's nontrivial topological properties under nonequilibrium conditions. 
When we continuously tune $\gamma$, will the changes in the topological property of NESS be different under different 
$\omega$ and $\ell$?  
We describe the influence of bond dissipation parameter $\gamma$ on the topological properties of the NESS via the location of the 
center-of-mass under different $\omega$ and $\ell$. In cold-atom experiments, the center-of-mass position can be 
determined by measuring the peak position of the Bose-Einstein condensate (BEC) \cite{BEC_1,BEC_2}. Similarly, 
in our numerical simulation, we use the probability center position of the wave function to characterize the particle's 
center-of-mass position. However, a key consideration 
arises: what is the critical distance, in units of the lattice constant, between the center-of-mass position and the system 
edge, below which the steady state can be considered an edge state? In cold-atom experiments, neutral atoms form a BEC 
at or below the critical temperature. This BEC is not perfectly localized at a single lattice site but exhibits a finite broadening, 
typically several times, or even tens of times, the lattice constant \cite{BEC_1,BEC_2}. This broadening implies that the BEC's 
center-of-mass position in the steady state is also located at least several lattice constants away from the system edge. 
Therefore, in numerical calculations, it is crucial to carefully determine the appropriate distance in lattice constants to 
definitively classify a steady state as an edge state.

\begin{figure}[htp]
		\centering
		\includegraphics[width=0.5\textwidth]{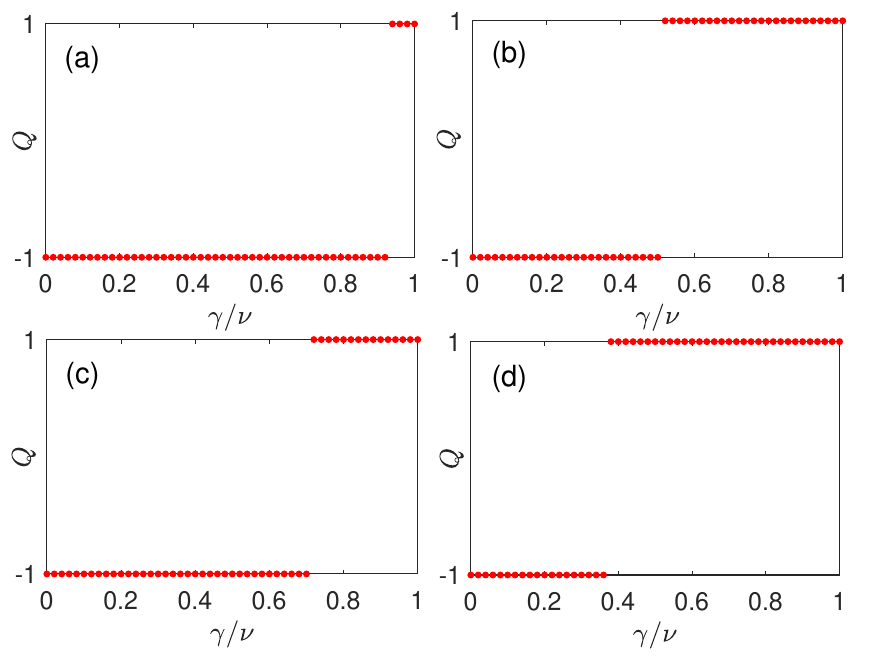}
		\caption{(Color online) Steady-state topological phase diagram as functions of $\gamma$.
		(a) $\ell=1$ and $\omega=1.5\nu$; (b) $\ell=1$ and $\omega=0.5\nu$; 
		(c) $\ell=2$ and $\omega=1.5\nu$; (d) $\ell=2$ and $\omega=0.5\nu$.
		Other parameter is $L=100$.
		}
		\label{f4}
\end{figure}

In experiments simulating Anderson localization, an optical well with trapping frequency $\omega \sim 10^{3}$ $ {\rm Hz}$ is
usually used to trap the neutral atoms, such as the $\prescript{39}{}{\rm K}$ or $\prescript{87}{}{\rm Rb}$ atoms.
When the system temperature is equivalent to or less than the BEC transition temperature, we can estimate the
broadening of the BEC formed by these cold atoms. This is because when the system temperature is at the transition
temperature of the BEC, the broadening of the BEC is approximately equal to the de Broglie wavelength ($\lambda_{\rm dB}$) \cite{BEC_1,BEC_2}.
Therefore, the broadening of BEC can be estimated by calculating the de Broglie wavelength. Having known the Planck constant $h$
and atomic mass $m$, $\lambda_{\rm dB}$ is estimated as $\lambda_{\rm dB} \sim h/\sqrt{2m\hbar\omega}$,
based on which, we estimate that de Broglie wavelength of $\prescript{39}{}{\rm K}$ is $\lambda^{\prescript{39}{}{\rm K}}_{\rm dB}=5.676~{\rm\mu m}$
and that of $\prescript{87}{}{\rm Rb}$ is $\lambda^{\prescript{87}{}{\rm Rb}}_{\rm dB}=3.79~{\rm \mu m}$. For the
Aubry-Andr\'{e}-type quasidisorder potential, Ref.~\cite{roati2008anderson} has proved that it is feasible to generate the quasiperiodically
optical wells by two standing waves with different frequencies, and the lattice constant is $516$ ${\rm nm}$ accordingly.
Meanwhile, the experimental reference also mentioned that the spatial size of $\prescript{39}{}{\rm K}$ condensate can be prepared
at $5$ ${\rm \mu m}$, so our estimated result is close to the experimental data \cite{roati2008anderson}. According to our estimated results,
The $\prescript{39}{}{\rm K}$ BEC is about $11$ times the lattice constant, and the $\prescript{87}{}{\rm Rb}$ BEC is about $7$-$8$ times
the lattice constant. It means that if the atoms are localized at the edges of system, the distance between the center-of-mass of atoms and
the edges of the system is about $3.5$-$5.5$ times the lattice constant.

Based on the above considerations, in our numerical calculation, we choose a relatively loose criterion, that is, 
the steady state where the distance between the center of mass and the system boundary is less than or equal to $5$ times 
the lattice constant is considered as the edge state. Analysis of Fig.~\ref{f4} reveals two critical dependencies: 
Firstly, topological phase transitions occur at significantly reduced critical values when the initial system is topologically trivial, 
as evidenced by comparing panels \ref{f4}(a) with \ref{f4}(b) or \ref{f4}(c) with \ref{f4}(d). Secondly, the parameter $\ell$ 
substantially modulates the transition threshold, with larger $\ell$ values lowering the critical point required for transition initiation, 
demonstrated through comparison of panels \ref{f4}(a) with \ref{f4}(c) or \ref{f4}(b) with \ref{f4}(d).

\begin{figure}[htp]
		\centering
		\includegraphics[width=0.5\textwidth]{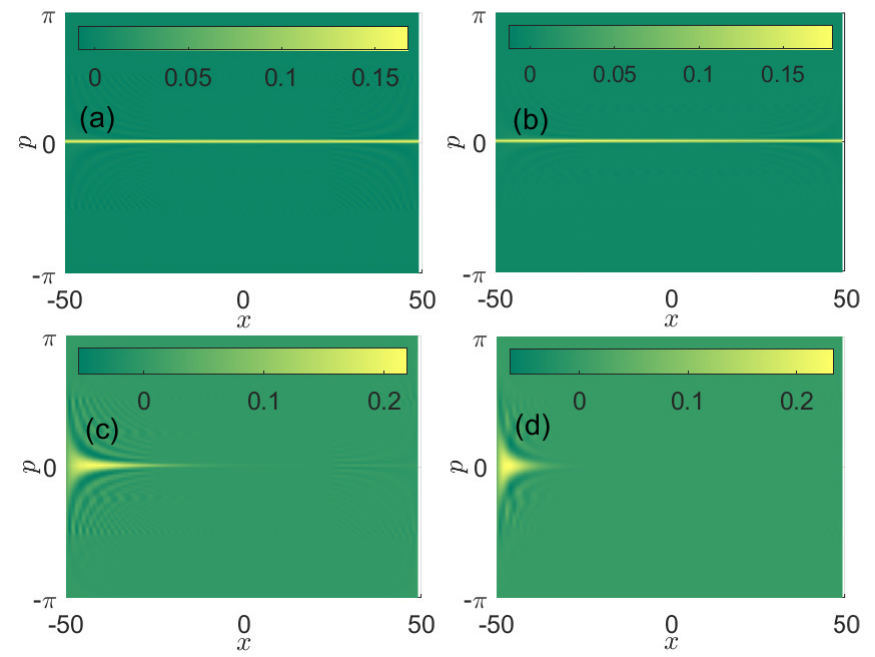}
		\caption{(Color online) Steady-state Wigner distributions.
		(a) $\gamma=0.1\nu$ and $\omega=1.5\nu$; (b) $\gamma=0.1\nu$ and $\omega=0.5\nu$;
		(c) $\gamma=\nu$ and $\omega=1.5\nu$; (d) $\gamma=\nu$ and $\omega=0.5\nu$.
		Other parameters are $\ell=1$ and $L=100$.
		}
		\label{f5}
\end{figure}

In addition to the spatial distribution of $\rho_{nn}$, 
the different forms of NESSs can be distinguished by using quantum phase space images. For example,
by solving the Wigner function in phase space, we can distinguish the cases with the edge state and the delocalized state. 
For a given quantum state, the Wigner distribution function is given by
\begin{equation}
W(x,p)=\frac{1}{2\pi\hbar}\int^{+\infty}_{-\infty} \phi^{*}(x-\frac{y}{2})\phi(x+\frac{y}{2})e^{-ipy/\hbar}dy,
\end{equation}
where $x$ and $p$ are the position and momentum in phase space, respectively, and $\hbar$ is the reduced Planck constant.
With the steady-state populations, we can construct the corresponding wave functions, i.e., $\ket{\psi}=\sum_{n}\phi(n)c^{\dag}_{n}\ket{0}$
with $\phi(n)=\sqrt{\rho_{nn}}$. Under $\ell=1$ and $\gamma=0.1\nu$, the steady-state Wigner distributions 
for $\omega=1.5\nu$ and $\omega=0.5\nu$ are plotted in Figs.~\ref{f5}(a) and \ref{f5}(b), respectively. 
We can see that Wigner distributions in the momentum branch
for both cases are localized at $p=0$. The distribution circumstances in the $x$ branch (at $p=0$) have analog to the steady-sate
populations (see Figs.~\ref{f2}(a) and \ref{f2}(b)). Under $\ell=1$ and $\gamma=\nu$, the Wigner distributions in the momentum
branch are delocalized at $x=-L/2$, and the distributions in the $x$ branch (at $p=0$) have analog to the steady-sate
populations (see Figs.~\ref{f2}(c) and \ref{f2}(d)).

\begin{figure}[htp]
		\centering
		\includegraphics[width=0.5\textwidth]{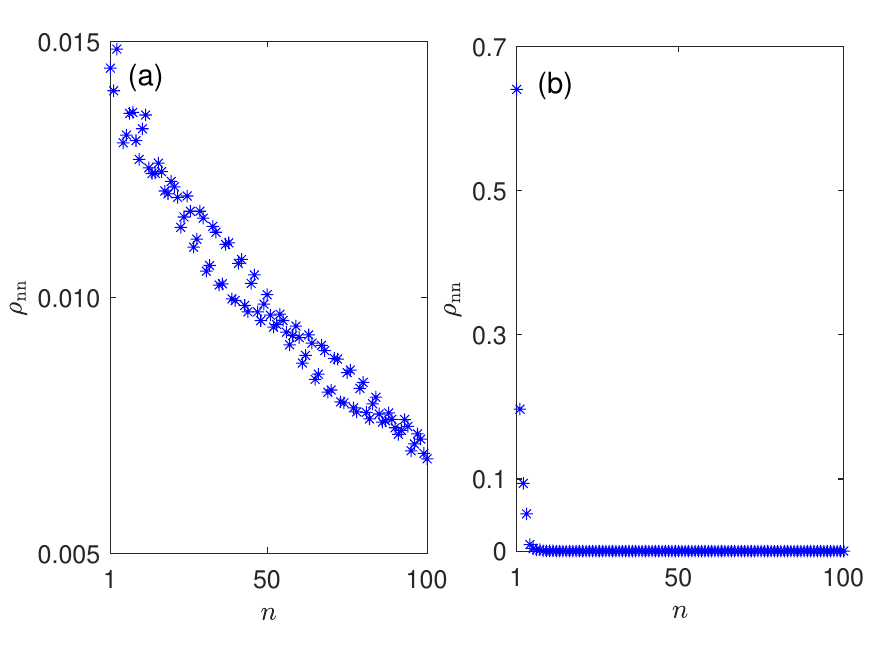}
		\caption{(Color online) Steady-state populations $\rho_{nn}$ as functions of site index under quasidisorder effect.  
		(a) $V=0.2\nu$; (b) $V=2\nu$.
		Other parameters are $\omega=0.5\nu$, $\ell=1$, $\gamma=0.1\nu$ and $L=100$.
		}
		\label{f6}
\end{figure}

\section{Quasidisorder Effect and Topological phase transition}\label{S3}
Reference \cite{Wang} reports that bond dissipation in a one-dimensional quasi-disordered system leads to 
delocalized or localized states, independent of the system's initial property. This prompts us to investigate whether 
similar or distinct behavior emerges when introducing a quasidisorder potential to the currently studied system. 
We consider an Aubry-André-type quasidisorder potential, $V_n=V \cos (2 \pi \alpha n) c_n^{\dagger} c_n$, 
where $\alpha=(\sqrt{5}-1) / 2$ is the incommensurate number. The total Hamiltonian is then ${H}_{t}=H+\sum_n V_n$. 
Previous studies \cite{Zhou_1,Zhou_2,Niaz} have shown that increasing $V$ drives a transition from a delocalized 
to a localized phase in the system described by ${H}_{t}$. We introduce bond dissipation to this Hamiltonian ${H}_{t}$ 
and examine the localization and topological properties of the resulting NESSs. Our findings reveal that the initial property of the 
system significantly influences the localization properties of the NESS after considering the quasidisorder effect as well. 
For instance, with $\ell=1, \omega=0.5 \nu$, 
and $V=0.2 \nu$, the system described by ${H}_{t}$ is initially in the delocalized phase. The steady-state 
population, shown in Fig.~\ref{f6}(a), still exhibits a delocalized state. Maintaining $\ell$ and $\omega$, but increasing 
$V$ to $ 2 \nu $, the corresponding steady-state populations, plotted in Fig.~\ref{f5}(b), demonstrate a edge localized
state. These results demonstrate that the initial property of the system has a profound impact on the localization properties of the NESS 
after considering the quasidisorder effect.

\begin{figure}[htp]
		\centering
		\includegraphics[width=0.5\textwidth]{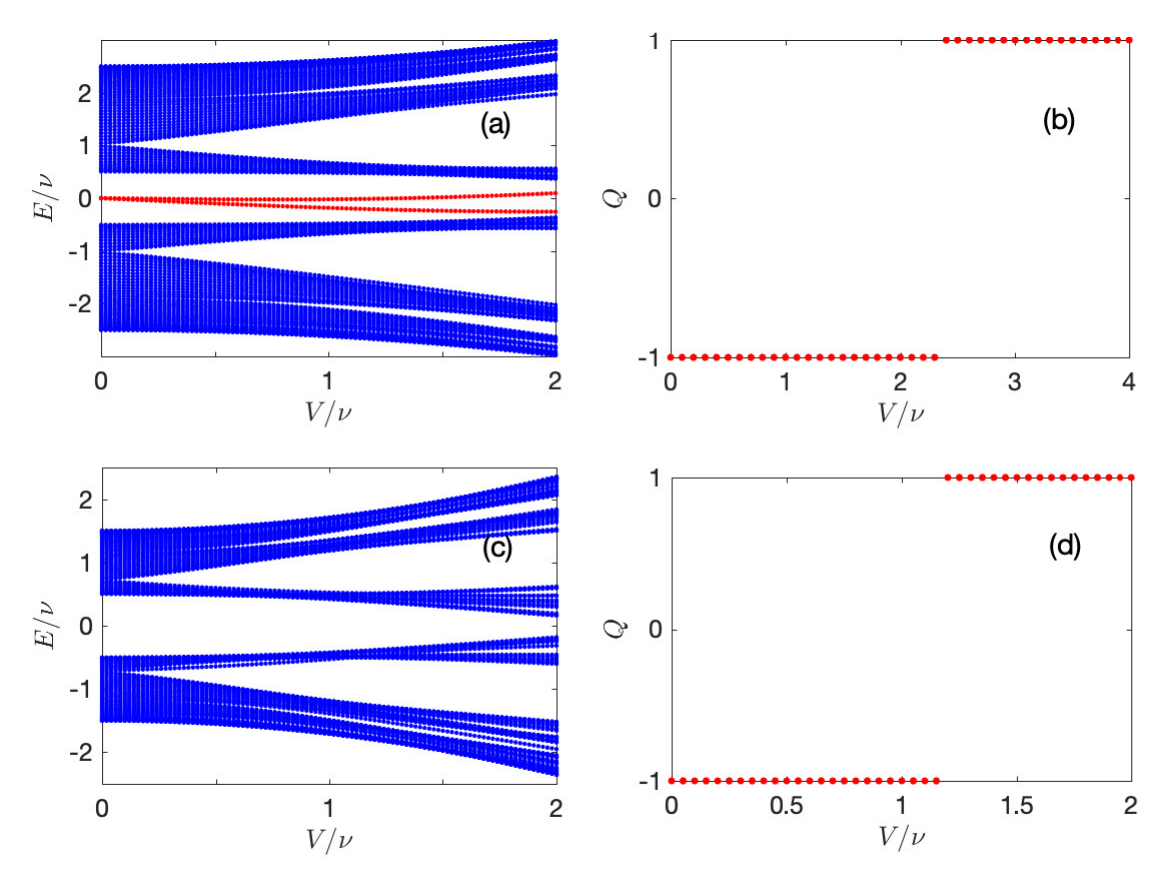}
		\caption{(Color online) (a) Energy Spectrum of $\mathcal{H}$ under $\omega=1.5\nu$.
		The red dots denote the edge states.
		(b) Topological phase diagram of steady-state under $\omega=1.5\nu$.
		 (c) Energy Spectrum of $\mathcal{H}$ under $\omega=0.5\nu$.
		(d) Topological phase diagram of steady-state under $\omega=0.5\nu$.
		Other parameters are $\ell=1$, $\gamma=0.1\nu$ and $L=100$.
		}
		\label{f7}
\end{figure}
The edge localization of the steady state reflects the system's nontrivial topological properties under 
nonequilibrium conditions. We describe the influence of a quasi-disordered potential on the topological 
properties of the NESS via the location of the center of mass as well. 
In our numerical calculation, we still choose a relatively loose criterion, that is, the steady state where the distance between the
center of mass and the system boundary is less than or equal to $5$ times the lattice constant is considered as the edge state.
Figure \ref{f7}(a) presents the energy spectrum of the initial system at $\omega=1.5\nu$. We can see that as the increase of
$V$, there are always two branches of edge states near zero energy, which are shown in red dots. It means that the initial system
has topological nontrivial property. As previously studied, a strong disorder potential will trigger the steady-state edge localization and
induce a steady-state topological phase transition. To intuitively describe the topological phase transition, we calculate the topological
phase diagram under $\ell=1$, $\gamma=0.1\nu$, and $\omega=1.5\nu$, and the phase diagram characterizing the existence of edge state is plotted in
Fig.~\ref{f7}(b). The parameter regions with $Q=1$ denote the existence of edge state, while those with
$Q=-1$ denote the absence of edge state (Here, $Q$ plays as a role of topological invariant. In fact, it only is a logic value
extracted from the distance.). We find that topological phase transitions emerge at substantial 
quasidisorder strengths. Crucially, for systems possessing initial topological non-triviality, 
the preservation of topological edge states in the NESS requires strong quasidisorder interventions.

Meanwhile, we calculate the initial energy spectrum for the $\omega=0.5\nu$ case, and it is plotted
in Fig.~\ref{f7}(c). As seen that, there is no any edge state in the spectrum.
We calculate the topological phase diagram for the $\omega=0.5\nu$ case in Fig.~\ref{f7}(d), and find that 
within the current region of quasidisorder strength, before a moderate quasidisorder strength, the NESS is still be a delocalized state. 
As the increase of $V$, exceeding the threshold,  the NESS experiences a topological phase transition
from the delocalized state to the edge state. Relative to the $\omega=1.5\nu$ case, the parameter regime 
exhibiting absence of edge states contracts significantly. It means that the initial topological trivial property 
will strengthen the topological non-triviality of the NESS. 
The intervention of a slightly moderate quasidisorder effect can make the NESS a topological edge state. 
Besides, the results show localization and topological properties of the NESS are greatly affected by the initial property of the system.

Having understood the localization and topological physics aroused by bond dissipation, now we briefly discuss the experimental feasibility.  
In 2013, Atala et al. formed a one-dimensional optical superlattice by superimposing 
two standing wave lasers of different frequency, and the expression of the lattice potential field is 
$V(x)=V_{1}\sin^{2}(k_{1}x+\phi)+V_{2}\sin^{2}(2k_{1}x+\pi/2)$ (where $V_{1(2)}$ denotes the strength of the potential, and $k_{1}$ and $2k_{1}$
denote the frequencies of the two lasers, respectively). Manipulating particle hopping between nearest neighboring sites by adjusting the phase $\phi$, 
the SSH model was realized \cite{SSH_exp}. By adjusting the frequencies of the two laser beams to non-synchrony parameters, 
Roati et al. constructed a quasi-periodic potential and observed the Anderson localization phenomenon \cite{roati2008anderson}. 
Therefore, with the aid of standing wave lasers and by adjusting parameters such as frequency and phase, our initial Hamiltonian 
without (with) quasidisordered potential can be experimentally realized. 

In addition, the bond dissipations with $\ell=1$ and $\ell=2$ can be realized by introducing the auxiliary lattice \cite{auxiliary_1,auxiliary_2,auxiliary_3,auxiliary_4,auxiliary_5}. 
For $\ell=1$ case, auxiliary sites are inserted between adjacent sites in the main lattice system (i.e., the initial lattice system), 
and the wavelength of the main lattice system matches the auxiliary lattice system. The coupling of the main lattice sites and 
the auxiliary sites is achieved through laser drive at antisymmetric Rabi frequencies $\pm \Omega$. By precisely controlling the 
emission process of atoms, the bond dissipation operator with $\ell=1$ can be realized \cite{auxiliary_1,auxiliary_2,auxiliary_3,auxiliary_4,auxiliary_5}. 
For $\ell=2$ case, it can be realized in a similar way. By using two hyperfine states, a spin-$\frac{1}{2}$ auxiliary lattice system is constructed, 
and the magnetic quantum numbers between the main and auxiliary lattice sites are conserved \cite{Wang}. Meanwhile, the wavelength of the laser which 
constructing the auxiliary lattice system is twice that of the main lattice system, so that it can leads to $\pi$ phase difference between the main 
and auxiliary lattice system. Specifically, the The odd (even) sites in the main lattice system correspond to the spin-down (spin-up) sites 
in the auxiliary lattice system \cite{Wang}. By manipulating the polarization of the driving laser, 
the hoppings between the main and auxiliary sites can be realized \cite{laser_1,laser_2,laser_3,laser_4}, thereby realizing the bond dissipation when $\ell=2$.
By detecting the transport behaviors \cite{tb_1,tb_2,tb_3,tb_4,tb_5,tb_6,tb_7} of atoms after introducing the bond dissipation, 
the bond dissipation induced delocalization-delocalization transition and topological phase transition can be observed.

\section{Summary}\label{S4}
We have investigated the localization and topological properties of the non-equilibrium steady state in a system subject to bond dissipation and quasi-disorder. Our findings contribute significantly to the understanding of dissipation-driven localization and topological phase transitions. Crucially, we demonstrate that the initial state of the system plays a critical role in determining the localization properties of the non-equilibrium steady state under bond dissipation. Analysis of steady-state populations and Wigner distributions reveals that dissipation-induced localization is not independent of the initial state; rather, it is strongly influenced by the initial system properties. One can engineer topological nontrivial systems by tuning the bond dissipation strength. Furthermore, we find that the combination of bond dissipation and quasi-disorder provides an effective method for controlling the topological properties of the system as well. 
This combination can drive transitions between topologically trivial and non-trivial states, and vice-versa.

This research is supported by Zhejiang Provincial Natural Science Foundation of China under Grant No. LQN25A040012, 
the National Natural Science Foundation of China under Grant No. 12174346, and the start-up fund from Xingzhi College, Zhejiang
Normal University.

\bibliography{ref}

\end{document}